\documentclass[a4paper]{jpconf}
\usepackage{graphicx}
\usepackage{wrapfig}
\usepackage{subfigure}

\begin{document}

\title{Azimuthal Anisotropy from  STAR Experiment}
\author{ Yadav Pandit (for the STAR Collaboration) }
\address{Department of Physics, University of Illinois at Chicago, USA}
\ead{ypandit@uic.edu}

\begin{abstract}
We report the measurement of  first (\textit{n} = 1) and higher order(\textit{n} = 2-5) harmonic coefficients ($v_{n}$)  of the azimuthal anisotropy  in the distribution of the particles produced  in  
Au+Au collisions at $\sqrt{s_{NN}}$ =200 GeV and U+U collisions at  $\sqrt{s_{NN}}$ =193 GeV, recorded  with the STAR detector at RHIC.  The differential measurement of $(v_{n})$  is  presented
 as a function of transverse momentum ($p_{T}$) and centrality. We also present  $v_{n}$  measurement  in the ultra-central collisions in U+U collisions.  These data may provide strong constraints on the theoretical models of the initial condition in heavy ion collisions and the transport properties of the produced medium.
\end{abstract}

\section{Introduction}
      
  The study of azimuthal anisotropy, based on Fourier coefficients, is widely recognized as an important tool to probe the hot, dense matter created in heavy ion collisions ~\cite{methodPaper}.  The distribution of particles  with respect to the participant planes can be written as
   
  \begin{equation}
 \frac {dN}  {d \Delta \phi}   \propto  1+  v_{1} \cos (\Delta \phi) +v_{2} \cos 2(\Delta \phi) +v_{3} \cos 3(\Delta \phi) + v_{4} \cos 4(\Delta \phi) +  . .     
  \label{eq1}
\end{equation}   
where $v_{n}$ is the $n^{th}$ order Fourier coefficient and  $\Delta \phi$ is the difference between $\phi$,  the azimuthal angle of each particle,  and  $\Psi_{n}$, the generalized participant event planes at all orders for each event.  The second harmonic also  called elliptic flow $v_{2}$  has been extensively studied both experimentally and theoretically.  Recently higher order harmonic also have gained considerable attention from theory~\cite{Mishra} and experimental community~\cite{v3Proceedings}.  These higher order harmonics can  provide  valuable information about the initial state of the colliding system~\cite{riseFall}  and also provide a natural explanation to the ridge phenomena in heavy ion collisions~\cite{derik}.  Measurement of these harmonics in different systems and collisions energy helps to further understand the heavy ion physics in general. 
 
 In uranium - uranium (U+U) collisions, there is  the potential to produce more extreme conditions of excited matter at higher density and/or greater volume than  is possible using spherical nuclei like lead or  less deformed nuclei like gold at the same incident energy~\cite{UUSim}. Uranium has quadrupole deformed shape. So U+U collisions may offer an opportunity to explore wider range of initial eccentricities.  The collisions of special interest are the ``ideal tip-tip" orientation in which the long axes of both deformed nuclei are aligned with the beam axis at zero impact parameter, and the ``ideal body-body" orientation in which the long axes are both perpendicular to the beam axis and parallel to each other at zero impact parameter. The ``ideal tip-tip" and ``ideal body-body" collision events allow to test the prediction of hydrodynamical models by varying the transverse particle density at spatial eccentricity similar to central Au+Au collisions.   
 
\section{The STAR experiment and Analysis Details}
 Data reported in this proceedings were collected in Au+Au collisions at $\sqrt{s_{NN}}$ = 200 GeV in the year 2004 with a minimum bias trigger and U+U collisions at $\sqrt{s_{NN}}$ = 193 GeV in the year 2012 with a minimum bias trigger  and the ultra central events  which were taken with a dedicated central trigger. The Time Projection Chamber (TPC)~\cite{startpc} is the primary tracking detector at STAR. Event  used in this analysis are required to have the primary vertex  along the beam direction  to be less than 30 cm and transverse direction to be less than  2 cm from the center of the beam pipe.  Centrality classes in U+U and Au+Au collisions are defined using the number  of charged particle tracks reconstructed in the TPC within pseudorapidity $|\eta| < 0.5 $ and passing within 3 cm of interaction vertex.       
  \begin{figure}[h]
\center
\includegraphics[width=35pc]{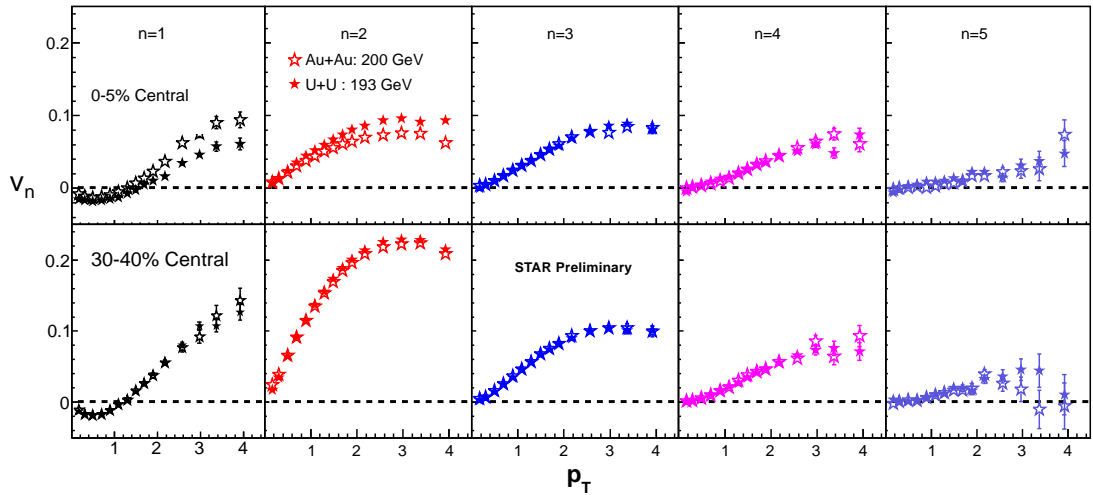}
\caption{$v_{n}(p_{T})$ measurement at 0-5\% central in upper panels and 30-40\% central collisions in the lower panels for U+U collisions at 193 GeV (solid stars) and for Au+Au collisions at 200 GeV(open stars).}
\label{fig1}
\end{figure}
 We used scalar product method as well as the event plane method to measure the signal~\cite{WWND2013}. In the scalar product method,  the subevents were  separated by  $\eta$  gap of 1.0 units between two subevents and at least 0.5 unit with the particle of interest and the subevent to which it is correlated .  In the event plane method,  the subevents were separated by  $\eta$   gap of 0.2 units.  Larger pseudorapidity separation was desired  but we are limited by the event plane resolution.  The event plane vector is reconstructed from tracks with transverse momentum ($p_{T}$) up to 2 GeV/$c$. This  suppresses the effects from high $p_{T}$ particles in the estimation of the event plane. The non-flow  contribution from the jets/minijets  is not known and might be a significant  contributor to the systematic uncertainties especially at peripheral collisions. 
 
\section{Result and discussion}

Results are presented with only statistical errors. In these studies, contribution  from short range correlation such as Bose-Einstein correlations, coulomb interactions  are studied using a pseudorapidity gap of at least 0.5 units in pseudorapidity between the event vector and the particle of interest using scalar product method. Results from both methods, scalar product with larger psuedorapidity  gap and   event plane method with smaller pseudorapidity gap  are consistent with each other, which suggests that non-flow contribution from short range correlations are small.

 \begin{figure}[h]
\center
\includegraphics[width=35pc]{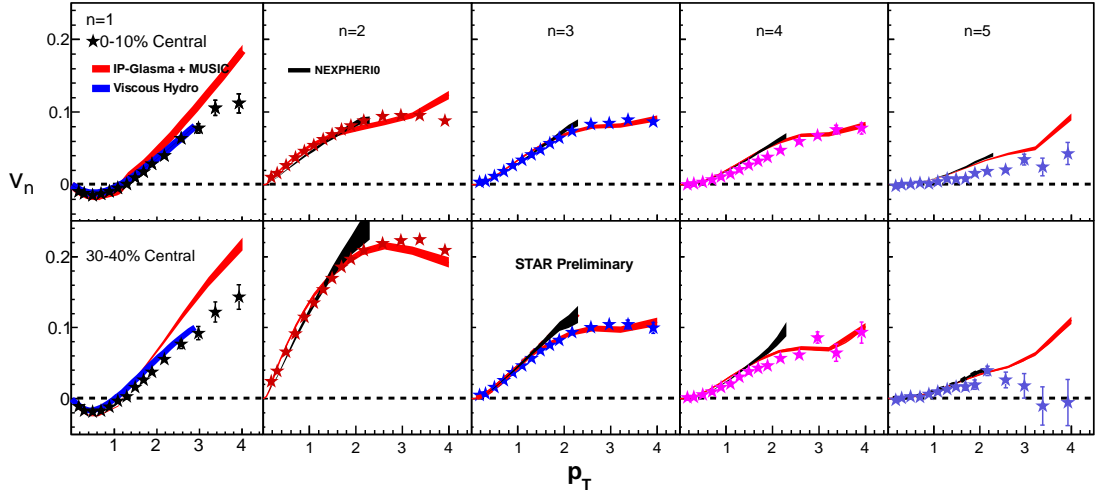}
\caption{$v_{n}$ measurement at 0-10\% central in upper panels and mid central (30-40\%) collisions in the lower panels as a function of $p_{T}$ for Au+Au collisions at 200 GeV compared with various model calculations}
\label{fig2}
\end{figure}
                                  
\begin{figure}
\begin{minipage}{0.37\linewidth}
\centerline{\includegraphics[width=1.2\linewidth]{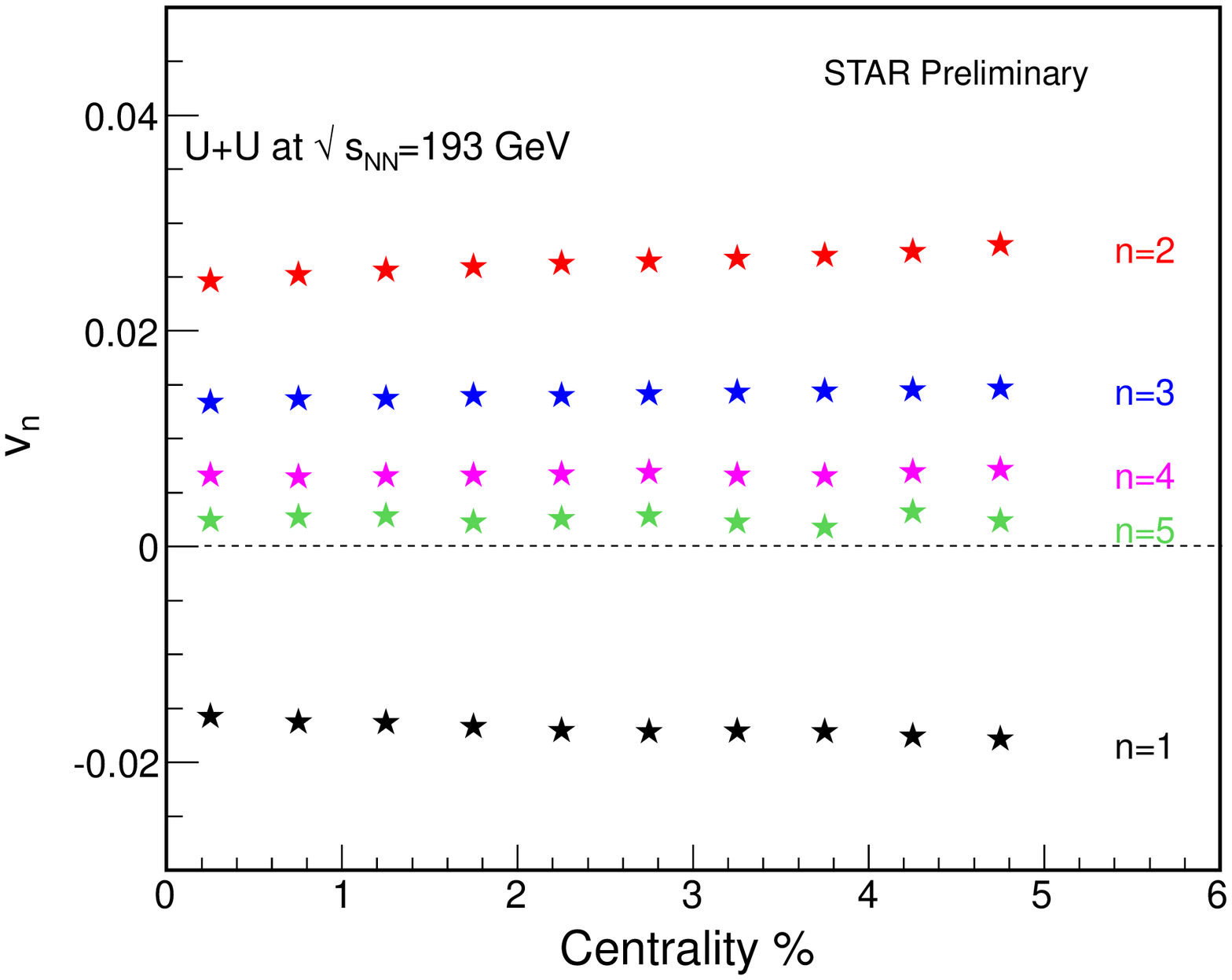}}
\end{minipage}
\hfill
\begin{minipage}{0.37\linewidth}
\centerline{\includegraphics[width=1.2\linewidth]{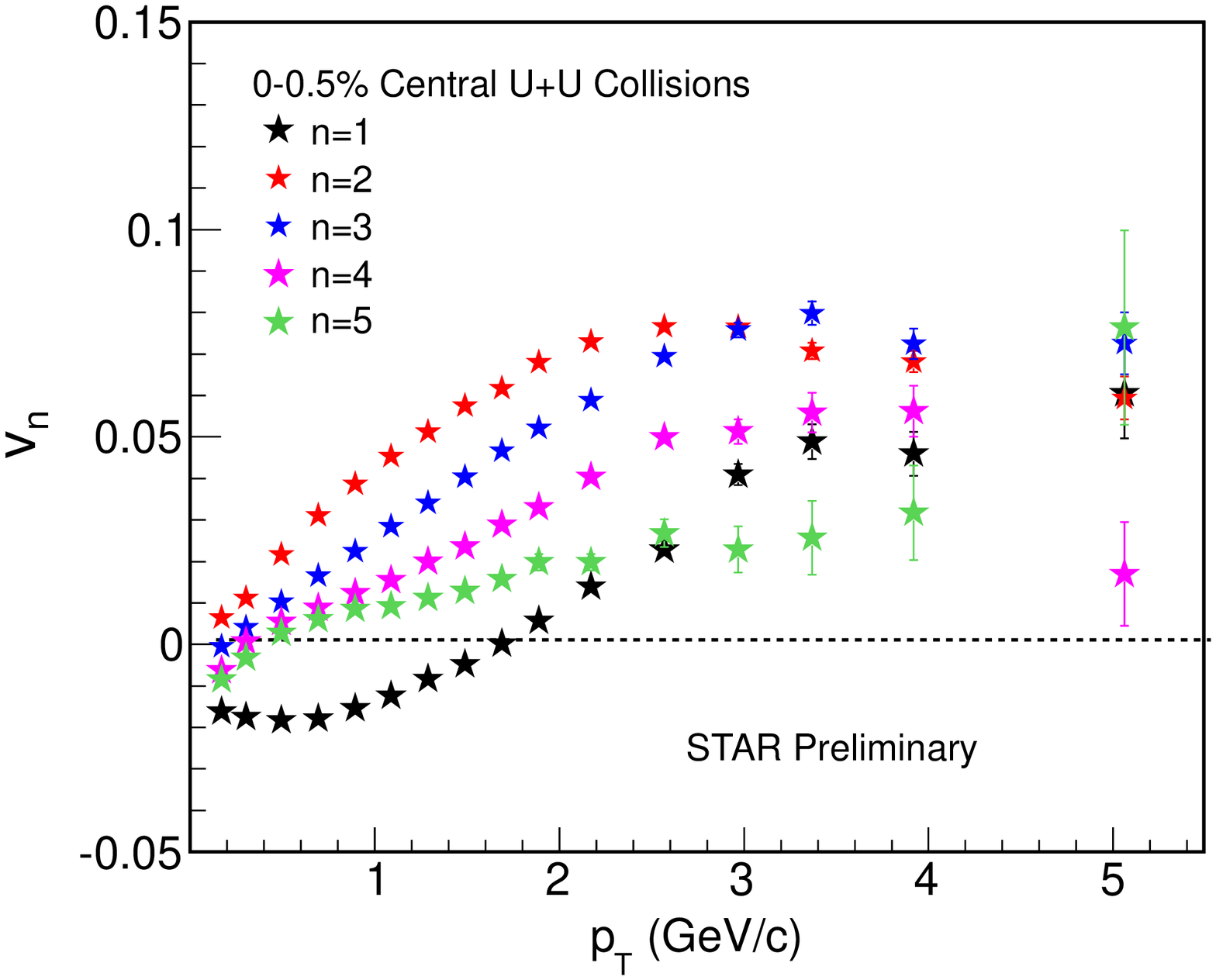}}
\end{minipage}
\caption{$v_{n}$ as a function of centrality at 0-5\% central collisions on left and  $v_{n}$ as a function of $p_{T}$ at ultra central(0-0.5\%) collisions on right }
\label{fig3}
\end{figure}

The $v_{n}$ of all charged hadrons, for $\textit{n} = 1, \, 2,\, 3,\, 4$ and 5  as a function of $p_{T}$ at various centralities, as shown in Fig.~\ref{fig1}.  The data for 0--5\% centrality are shown in the upper panels, and for intermediate centrality (30--40\%) in the lower panels.  All $v_{n}$ measurements show an increasing trend as a function of $p_{T}$.  Except for $v_{2}$, we observe very weak centrality dependence. Also shown are the same measurement  from Au+Au collisions at 200 GeV for the comparisons. We observe the difference in $v_{n}$ for  \textit{n} = 1 and \textit{n} = 2 at 0-5\% central between U+U
and Au+Au collisions. This may hint  to  initial overlap geometry difference in the central collisions between two systems Au+Au and U+U collisions. The difference diminishes in higher harmonics and more peripheral collisions.        

Fig.~\ref{fig2}  shows the flow harmonics $v_{n}$ for $n=1$ to $n=5$ as a function of $p_{T}$  in 200 GeV Au+Au collisions compared to the model calculations. In the ideal hydrodynamic  NeXSpheRIO  calculation ~\cite{NEXus}(shown in black line), the fluctuating initial condition is introduced from MC event generator Nexus. The IP$\_$Glasma $+$ Music ~\cite{IPGlasma} viscous hydrodynamic calculation(shown in red line) uses IP$-$Glasma initial condition and low viscosity setting of $\eta/s$ =0.12. For $v_{1}$, we also compare with prediction from viscous hydrodynamic extrapolation from LHC energy to RHIC energy~\cite{VHydro}(shown in blue line),  they use a smooth density profile with an added deformation to reproduce the data.  For lower harmonics and central collisions, the models with low viscosity setting and fluctuating initial condition or added deformation to the smooth initial density profile describe the data well however the model curve overshoot in the higher harmonics and peripheral collisions at high $p_{T}$

We also report measurement of $v_{n}$ for 0-5\% central U+U collisions taken with a dedicated central trigger. We subdivide 0-5\% central bin into 10 smaller centrality bins upto most central 0-0.5\% centrality.   In Fig.~\ref{fig3} (a), $v_{n}$ as a function of centrality for ultra central collisions is shown. Other than second harmonic coefficient, the $v_{n}$  do not change in this centrality range. We observe small change for $v_{2}$ since it still has some contribution from initial overlap geometry. This observation suggests that there is still a small contribution from body-body collisions. 
  In Fig.~\ref{fig3} (b), the $v_{1}$ signal along with $v_{2}$, $v_{3}$,  $v_{4}$ and  $v_{5}$ a function of transverse momentum up to $p_{T}$  $\sim$ 5 GeV/$c$  is shown for ultra central (0-0.5\%) collisions. The most central  0-0.5\%  collisions is expected to have dominant contribution from tip-tip collisions.  In these most central events we find that higher harmonics are also significant in magnitude compared with  second harmonics in the intermediate transverse momentum. Contributions  from higher harmonics should not be overlooked interpreting  dihadrom correlation data at intermediate transverse momentum. These higher harmonics may offer natural explanation to the novel ridge phenomena in heavy ion collisions~\cite{riseFall}.   Model comparisons with these new data may help us to better understand the medium properties. 
  
\section {Summary}
 We report the first  measurement of  azimuthal anisotropy  $v_{n}$  for  \textit{n}=1-5  as a function of transverse momentum $p_{T}$ and centrality in U+U collisions at  $\sqrt{s_{NN}}$ =193 GeV, recorded  with the STAR detector at RHIC.  Centrality dependence is weak for harmonics other than second harmonics. For higher harmonics and mid central collisions, $v_{n}$(U+U) is similar to $v_{n}$(Au+Au),  the difference appears at central collisions for $v_{1}$ and $v_{2}$.  At  intermediate $p_{T}$  range 3-5 GeV/$c$,  $v_{n}$'s are  comparable to the $v_{2}$ signal in ultra  central collisions. Model calculation specially at ultra central collisions  may be useful to constrain the initial condition and  transport coefficient of the medium produced in heavy ion collisions.

\end{document}